\documentclass[12pt, prd, showpacs]{revtex4}
\usepackage{amssymb}
\usepackage{amsmath}

\input{tcilatex}

\begin{document}

\title{Innermost stable circular orbit near dirty black holes in magnetic
field and ultra-high energy particle collisions}
\author{O. B. Zaslavskii}
\email{zaslav@ukr.net }
\affiliation{Department of Physics and Technology, Kharkov V.N. Karazin National
University, 4 Svoboda Square, Kharkov 61022, Ukraine}
\affiliation{Institute of Mathematics and Mechanics, Kazan Federal University, 18
Kremlyovskaya St., Kazan 420008, Russia}

\begin{abstract}
We consider the behavior of the innermost stable circular orbit (ISCO) in
the magnetic field near "dirty" (surrounded by matter) axially-symmetric
black holes. The cases of near-extremal, extremal and nonextremal black
holes are analyzed. For nonrotating black holes, in the strong magnetic
field ISCO approaches the horizon (when backreaction of the field on
geometry is neglected). Rotation destroys this phenomenon. The angular
momentum and radius of ISCO look model-independent in the main
approximation. We also study the collisions between two particles that
results in the ultra-high energy $E_{c.m.}$ in the centre of mass frame. Two
scenarios are considered - when one particle moves on the near-horizon ISCO
or when collision occurs on the horizon, one particle having the energy and
angular momentum typical of ISCO. If the magnetic field is strong enough and
a black hole is slow rotating, $E_{c.m.}$ can become arbitrarily large.
Kinematics of high-energy collision is discussed. As an example, we consider
the magnetized Schwarzschild black hole for an arbitrary strength of the
field (the Ernst solution). It is shown that backreaction of the magnetic
field on the geometry can bound the growth of $E_{c.m.}$.
\end{abstract}

\maketitle

\section{Introduction}

Motion of particles in the vicinity of black holes is a subject that has
been continuing to attract interest until recently. In doing so, a special
role is played by circular orbits - see. e.g., recent papers \cite{rufk}, 
\cite{ruf3} and references therein. Especially, this concerns an innermost
stable circular orbit (ISCO). It is important in phenomena connected with
accretion disc and properties of cosmic plasma \cite{ac1}, \cite{ac2}. Apart
from astrophysics, such a kind of orbits posseses a number of nontrivial
features and, therefore, is interesting from the theoretical viewpoint. In a
"classic" \ paper \cite{72} it was shown that in the extremal limit ISCO
approaches the horizon. As a result, some subtleties arise here since the
horizon is a ligntlike surface, so a massive particle cannot lie within it
exactly. Nowadays, near-horizon circular orbits for near-extremal and
extremal rotating black holes is still a subject of debates \cite{ted11}, 
\cite{ind2}, \cite{m}, \cite{extc}.

Quite recently, a new circumstance came into play that makes the properties
of ISCO important in a new context. Namely, it is the ISCO that turns out a
natural venue for realizaiton of the so-called BSW\ effect. Several years
ago, it was shown by Ba\~{n}ados, Silk and West that if two particles
collide near the black hole horizon of the extremal Kerr metric, their
energy $E_{c.m.}$ in the centre of mass (CM) frame can grow unbound \cite%
{ban}. This is called the BSW effect, after the names of its authors. These
findings stimulated further study of high-energy collisions near black
holes. The validity of the BSW effect was extended to extremal and
nonextremal more general black holes. It was also found that there exists
the version of this effect near nonrotating electrically charged black holes 
\cite{jl}. Another version of ultra-high energy collisions reveals itself in
the magnetic field, even if a black hole is neutral, vacuum and nonrotating,
so it is described by the Schwarzschild metric \cite{fr}. Generalization to
the case when the background is described by the Kerr metric was done in 
\cite{weak}.

In the BSW effect, one of colliding particle should be so-called critical.
It means that the energy and the angular momentum (or electric charge) of
this particle should be fine-tuned. In particular, the corresponding
critical condition is realized with good accuracy if a particle moves on a
circular orbit close to the horizon. Therefore, an innermost stable circular
orbit (ISCO) can play a special role in ultra-high energy collisions in
astrophysical conditions. Without the magnetic field, this was considered in 
\cite{circkerr} for the Kerr black hole and in \cite{circ} for more general
rotating black holes. Kinematically, the effect is achieved due to collision
of a rapid typical so-called usual particle (without fine-tuning) and the
slow fine-tuned particle on the ISCO \cite{k} (see also below).

In \cite{fr} and \cite{weak} collisions were studied just near the ISCO in
the magnetic field. In both cases, a black hole was taken to be a vacuum
one. Meanwhile, in astrophysical conditions, black holes are surrounded by
matter. By definition, such black holes are called "dirty", according to the
terminolgy suggested in Ref. \cite{vis}. (We wold like to stress that it is
matter but not the electromagnetic field that makes a black hole dirty.)

The aim of our work is two-fold since two different issues overlap here. The
first one is the properties of ISCO near dirty black holes in the magnetic
field, so both matter and the magnetic field are present. The second issue
is the scenarios of high-energy particle collisions near such orbits. We
derive general asymtotic formulas for the poistion of the ISCO in the
magnetic field which are used further for the evaluaiton of $E_{c.m.}$ and
examining of two scenarios of the BSW effect near ISCO.

In both works \cite{fr} and \cite{weak}, it was assumed that the magnetic
field is weak in the sense that backreaction of the magnetic field on the
metric is negligible but, at the same time, it is strong in the sense that
it affects motion of test particles. Such combination is self-consistent
since the dimensionless parameter $b$ that controls the magnetic field
strength contains a large factor $q/m$ relevant for motion of particles. Our
approach is model-independent and is not restricted by some explicit
background metric.\ Therefore, the most part of formulas applies also to the
metrics which are affected by the magnetic field. From the other hand, if
the magnetic field is too strong, its backreaction on the metric can change
the properties of $E_{c.m.}$ itself, as will be seen below. Thus we discuss
two news features absent from previous works in the sense that both matter
and magnetic field are taken into account in a model-independent way.

It is worth noting that high-energy collisions in the magnetic field were
studied also in another context, including scenarions not connected with
ISCO - see \cite{string} - \cite{tur2}.

In general, it is hard to find and analyze ISCO even in the Kerr or
Kerr-Newman cases \cite{rufk}, \cite{ruf3}. However, it is the proximity to
the horizon that enables us to describe some properties of ISCO, even not
specifying metric (so in a model-independent way) and even with the magnetic
field.\ This can be considered as one of manifestation of universality of
black hole physics.

The paper is organized as follows. In Sec. II, the metric and equations of
motion are presented. In Section III, we give basic equations that determine
ISCO. In Sec. IV, we consider ISCO in the magnetic field for near-extremal
black holes and analyze the cases of small and large fields. In Sec. V the
case of nonrotating (but dirty) and slowly rotating black hole is discussed.
As we have two small parameters (slow rotation and inverse field strength),
we consider different relations between them separately. In Sec. VI we show
that if a black hole rotates, even in the limit of strong magnetic field
ISCO does not tend to the horizon radius. In Sec. VII, it is shown that for
extremal nonrotating black holes, for large $b$, ISCO approaches the horizon
radius. In Sec. VIII, it is shown that this property is destroyed by
rotation. In Sec. IX, general formulas for $E_{c.m.}$ for particle
collisions in the magnetic field are given. In Sec. X, we find the velocity
of a particle on ISCO and argue that kinematic explanation of high-energy
collision is similar to that for the BSW effect \cite{k}. In Sec. XI, we
apply general formulas of collision to different black hole configurations
and different scenarios. In Sec. XII, the exact solution of the
Einstein-Maxwell equations (static Ernst black hole) is chosen as background
for collisions. This enables us to evaluate the role of backreaction of the
magnetic field on $E_{c.m.}$. In Sec. XIII, the main results are summarized.
Some technical points connected with cumbersome formulas are put in Appendix.

Throughout the paper we use units in which fundamental constants are $G=c=1$.

\section{Metric and equations of motion}

Let us consider the metric of the form 
\begin{equation}
ds^{2}=-N^{2}dt^{2}+\frac{dr^{2}}{A}+R^{2}(d\phi -\omega dt)^{2}+g_{\theta
}d\theta ^{2}\text{,}
\end{equation}%
where the metric coefficients do not depend on $t$ and $\phi $. The horizon
corresponds to $N=0$. We also assume that there is an electromagnetic field
with the four-vector $A^{\mu }$ where the only nonvanishing component equals 
\begin{equation}
A^{\phi }=\frac{B}{2}\text{.}  \label{ab}
\end{equation}

In vacuum, this is an exact solution with $B=const$ \cite{wald}. We consider
configuration with matter (in this sense a black hole is "dirty"), so in
general $B$ may depend on $r$ and $\theta $.

Let us consider motion of test particles in this background. The kinematic
momentum $p^{\mu }=mu^{\mu }$, where $m$ is the particle's mass,
four-velocity $u^{\mu }=\frac{dx^{\mu }}{d\tau }$, where $\tau $ is the
proper time, $x^{\mu }$ are coordinates. Then, the generalized momentum is
equal to

\begin{equation}
p_{\mu }=P_{\mu }-qA_{\mu }\text{,}
\end{equation}%
$q$ is the particle's electric charge. Due to the symmetry of the metric, $%
P_{0}=-E$ and $P_{\phi }=L$ are conserved, where $E$ is the energy, $L$ is
the angular momentum.

We consider motion constrained within the equatorial plane, so $\theta =%
\frac{\pi }{2}$. Redefining the radial coordinate $r\rightarrow \rho $, we
can always achieve that 
\begin{equation}
A=N^{2}  \label{an}
\end{equation}%
within this plane. Then, equations of motion give us 
\begin{equation}
\dot{t}=\frac{X}{N^{2}m}\text{,}  \label{t}
\end{equation}%
\begin{equation}
\dot{\phi}=\frac{\beta }{R}+\frac{\omega X}{mN^{2}}\text{,}  \label{ft}
\end{equation}%
\begin{equation}
m^{2}\dot{\rho}^{2}+V=0\text{,}  \label{rt}
\end{equation}%
\begin{equation}
X=E-\omega L\text{,}  \label{X}
\end{equation}%
\begin{equation}
\beta =\frac{\mathcal{L}}{R}-\frac{qBR}{2m}\text{,}  \label{b}
\end{equation}%
\begin{equation}
V=m^{2}N^{2}(1+\beta ^{2})-X^{2}\text{.}  \label{v}
\end{equation}%
Dot denotes differentiation with respect to the proper time $\tau $. As
usual, we assume the forward in time condition $\dot{t}>0$, so $X\geq 0$.
Hereafter, we use notations

\begin{equation}
\mathcal{L\equiv }\frac{L}{m}\text{, }\mathcal{E}=\frac{E}{m},\text{ }b=%
\frac{qB_{+}R_{+}}{2m}\text{.}  \label{bb}
\end{equation}%
Subscripts "+", "0" denote quantities calculated on the horizon and ISCO,
respectively.

In what follows, we will use the Taylor expansion of quantity $\omega $ near
the horizon. We denote $x=\rho -\rho _{+}$, where $\rho _{+}$ is the horizon
radius. Then,

\begin{equation}
\omega =\omega _{+}-a_{1}x+a_{2}x^{2}+...  \label{om}
\end{equation}

\section{Equations determining ISCO}

By definition, ISCO is determined by equations \cite{72}%
\begin{equation}
V(\rho _{0})=0\text{,}  \label{0}
\end{equation}%
\begin{equation}
\frac{dV}{d\rho }(\rho _{0})=0\text{,}  \label{fd}
\end{equation}%
\begin{equation}
\frac{d^{2}V}{d\rho ^{2}}(\rho _{0})=0.  \label{sd}
\end{equation}

Eqs. (\ref{v}), (\ref{0}) entail%
\begin{equation}
X(\rho _{0})=mN(\rho _{0})\sqrt{1+\beta ^{2}(\rho _{0})}  \label{xx}
\end{equation}%
and eqs. (\ref{fd}), (\ref{sd}) turn into%
\begin{equation}
\frac{1}{m^{2}}\frac{dV_{eff}}{d\rho }=\frac{d}{d\rho }[N^{2}(1+\beta
^{2})]+2\mathcal{L}\omega ^{\prime }\sqrt{1+\beta ^{2}}N=0\text{,}
\label{1d}
\end{equation}%
\begin{equation}
\frac{1}{m^{2}}\frac{d^{2}V_{eff}}{d\rho ^{2}}=\frac{d^{2}}{d\rho ^{2}}%
[N^{2}(1+\beta ^{2})]-2\mathcal{L}^{2}\omega ^{\prime }\sqrt{1+\beta ^{2}}%
\omega ^{\prime 2}+2\mathcal{L}\omega ^{\prime }\sqrt{1+\beta ^{2}}\omega
^{\prime \prime }N\sqrt{1+\beta ^{2}}=0,  \label{2d}
\end{equation}%
where all quantities in (\ref{1d}), (\ref{2d}) are to be taken at $\rho
=\rho _{0}$. Prime denotes derivative with respect to $\rho $ (or,
equivalently, $x$).

In general, it is impossible to find exact solutions of eqs. (\ref{1d}), (%
\ref{2d}). Therefore, in next sections we analyze separately different
particular situations, with main emphasis made on the near-horizon region.
In doing so, we develop different versions of the perturbation theory that
generalize the ones of \cite{weak}. The radius of ISCO, its energy and
angular momentum are represented as some series with respect to the
corresponding small parameter, truncated at the leading or subleading terms
similarly to \cite{weak}.

\section{Near-extremal black holes}

Let a black hole be nonextremal. In what follows, we are interested in the
immediate vicinity of the horizon and use the Taylor series for
corresponding quantities. Then, near the horizon we have the expansion

\begin{equation}
N^{2}=2\kappa x+Dx^{2}+Cx^{3}...\text{,}  \label{N}
\end{equation}%
where $\kappa $ has the meaning of the surface gravity.

By definition, we call a black hole near-extremal if%
\begin{equation}
\kappa \ll Dx_{0}\text{,}
\end{equation}%
where $x_{0}=\rho _{0}-\rho _{+}$. Then, for the lapse function we have
expansion near ISCO%
\begin{equation}
N=x\sqrt{D}+\frac{\kappa }{\sqrt{D}}-\frac{\kappa ^{2}}{2D^{3/2}x}+\frac{C}{2%
\sqrt{D}}x^{2}+...  \label{nx}
\end{equation}

Taking into account (\ref{2d}), after straightforward (but somewhat
cumbersome) calculations, one can find that%
\begin{equation}
-\frac{1}{2}\frac{dV_{eff}}{d\rho }(\rho _{0})=A_{2}x^{2}+A_{3}\frac{\kappa
^{2}}{x}+...  \label{1}
\end{equation}%
\begin{equation}
\mathcal{L}a_{1}\approx \sqrt{DP}\text{,}  \label{ld}
\end{equation}%
\begin{equation}
P\equiv 1+\beta ^{2}\text{,}  \label{p}
\end{equation}%
\begin{equation}
A_{2}\approx \frac{D}{2}\frac{dP}{dx}+\frac{CP}{2}+\frac{a_{2}}{a_{1}}PD
\end{equation}

\begin{equation}
A_{3}=-\frac{P}{2D},
\end{equation}%
where $P$ and $\frac{dP}{dx}$ are to be taken at $x=x_{0}$ or, with the same
accuracy, at $x=0$ (i.e., on the horizon).

Then, 
\begin{equation}
x_{0}^{3}\approx -\frac{A_{3}}{A_{2}}\kappa ^{2}=H^{3}\kappa ^{2}\text{,}
\label{h}
\end{equation}%
\begin{equation}
H=(\frac{P_{0}}{2DA_{2}})^{1/3}=\frac{1}{[D(2\frac{a_{2}}{a_{1}}D+C+\frac{D}{%
P}\frac{dP}{dx})]^{1/3}}\text{.}  \label{H}
\end{equation}

From (\ref{xx}), (\ref{nx}) we have 
\begin{equation}
N_{0}\approx \sqrt{D}H\kappa ^{2/3}\text{,}  \label{nk}
\end{equation}%
\begin{equation}
X_{0}\approx m\sqrt{P_{+}}\sqrt{D}H\kappa ^{2/3}\text{.}  \label{x23}
\end{equation}

Using (\ref{ld}), (\ref{p}) and (\ref{b}) we derive equation for the value
of the angular momentum $L_{0}$ on ISCO:

\begin{equation}
\frac{\mathcal{L}_{0}^{2}}{R_{+}^{2}}+2b\frac{D}{d-D}\frac{\mathcal{L}_{0}}{%
R_{+}}-\frac{D(1+b^{2})}{d-D}=0.
\end{equation}%
where 
\begin{equation}
d\equiv R_{+}^{2}a_{1}^{2}.  \label{da}
\end{equation}%
To have a well-defined limit $b=0$, we demand $d-D>0$. We are interested in
the positive root according to (\ref{ld}). Then, 
\begin{equation}
\frac{\mathcal{L}_{0}(b)}{R_{+}}=-\frac{bD}{d-D}+\frac{\sqrt{D}}{d-D}\sqrt{%
d(1+b^{2})-D},  \label{l0}
\end{equation}%
and, in a given approximation,%
\begin{equation}
\beta _{0}=\frac{1}{d-D}[\sqrt{D}\sqrt{d(1+b^{2})-D}-bd]\text{, }
\end{equation}%
\begin{equation}
P_{0}=\frac{d}{d-D}-2\frac{b\sqrt{D}d}{(d-D)^{2}}\sqrt{d(1+b^{2})-D}+\frac{%
b^{2}d(d+D)}{(d-D)^{2}}\text{,}
\end{equation}%
\begin{equation}
\left( \frac{d\beta }{dx}\right) _{+}=-\frac{R_{+}^{\prime }}{R_{+}}[b+\frac{%
\mathcal{L}_{0}(b)}{R_{+}}]-\frac{B_{+}^{\prime }}{B_{+}}b\text{,}
\end{equation}%
\begin{equation}
A_{2}\approx D\beta _{0}\left( \frac{d\beta }{dx}\right) _{+}+\frac{CP_{0}}{2%
}+\frac{a_{2}}{a_{1}}P_{0}D\text{,}  \label{a2}
\end{equation}%
where we neglected the difference between $\left( \frac{d\beta }{dx}\right)
_{+}$ and $\left( \frac{d\beta }{dx}\right) _{0}$. Eqs. (\ref{l0}) - (\ref%
{a2}) give the expression for $H$ after substitution into (\ref{H}). To
avoid cumbersome expressions, we leave it in the implicit form.

Now, two different limiting cases can be considered.

\subsection{Small magnetic field}

If $B=0$, 
\begin{equation}
\mathcal{L}_{0}(0)=\frac{\sqrt{D}}{\sqrt{a_{1}^{2}-\frac{D}{R_{+}^{2}}}},
\end{equation}%
\begin{equation}
\beta (0)=\frac{\sqrt{D}}{\sqrt{d-D}}\text{,}  \label{bsm}
\end{equation}%
\begin{equation}
P_{0}(0)\approx \frac{d}{d-D}=\frac{R_{+}^{2}a_{1}^{2}}{R_{+}^{2}a_{1}^{2}-D}
\end{equation}%
that agrees with eq. (44) of Ref. \cite{circ}. It follows from (\ref{xx})
and (\ref{x23}) that 
\begin{equation}
\mathcal{E}(0)\approx \omega _{+}\frac{\sqrt{D}}{\sqrt{a_{1}^{2}-\frac{D}{%
R_{+}^{2}}}}\text{.}
\end{equation}%
Let us consider small but nonzero $b$. We can find from (\ref{l0}) that 
\begin{equation}
\mathcal{L}_{0}(b)\approx \mathcal{L}_{0}(0)-\mathcal{L}_{0}^{2}(0)\frac{b}{%
R_{+}}+O(b^{2}),
\end{equation}%
\begin{equation}
\mathcal{E}_{0}(b)\approx \omega _{0}\mathcal{L}_{0}+O(\kappa ^{2/3}\text{, }%
b^{2})\text{.}
\end{equation}

\subsection{Large magnetic field}

Let $b\gg 1$. Now, $P_{0}\sim b^{2}$, $A_{2}\sim b^{2}.$ According to (\ref%
{h}), there exists a finite $\lim_{B\rightarrow \infty }H=H_{\infty }$. In
doing so, we find from (\ref{l0}), (\ref{b}), (\ref{x23}) 
\begin{equation}
\frac{\mathcal{L}_{0}}{R_{+}}\approx \frac{b\sqrt{D}}{\sqrt{d}+\sqrt{D}}=%
\frac{b\sqrt{D}}{a_{1}R_{+}+\sqrt{D}}\text{,}  \label{lal}
\end{equation}%
\begin{equation}
\beta \approx -\frac{\sqrt{d}b}{\sqrt{d}+\sqrt{D}}=-\frac{R_{+}a_{1}b}{%
R_{+}a_{1}+\sqrt{D}}\text{,}  \label{bs}
\end{equation}%
\begin{equation}
P_{0}\approx b^{2}\frac{d}{(\sqrt{d}+\sqrt{D})^{2}},
\end{equation}%
\begin{equation}
X_{0}\approx m\sqrt{D}H_{\infty }\kappa ^{2/3}\frac{ba_{1}R_{+}}{a_{1}R_{+}+%
\sqrt{D}}\text{,}
\end{equation}%
\begin{equation}
\mathcal{E}\approx b\sqrt{D}[\frac{\omega _{+}R_{+}}{a_{1}R_{+}+\sqrt{D}}+%
\sqrt{D}H_{\infty }\kappa ^{2/3}\frac{a_{1}R_{+}}{a_{1}R_{+}+\sqrt{D}}]\text{%
.}  \label{el}
\end{equation}

Thus according to (\ref{h}), in general the radius of ISCO depends on the
value of the magnetic field via the coefficient $H$. However, there is an
exception. Let 
\begin{equation}
C=0\text{, }R_{+}^{\prime }=0\text{, }B^{\prime }=0.  \label{c}
\end{equation}

Then, 
\begin{equation}
H^{3}=-\frac{A_{3}}{A_{2}}=\frac{a_{2}}{2a_{1}}\frac{1}{D^{2}}\text{,}
\end{equation}%
so the dependence on $b$ drops out from the quantity $H$ and,
correspondingly, from the ISCO radius (\ref{h}). One can check easily that
the conditions (\ref{c}) are satisfied for the near-extremal Kerr metric in
the magnetic field. This agrees with eq. (38) of \cite{weak} where the
observation was made that in the main corrections of the order $\kappa
^{2/3} $ the magnetic field does not show up. Thus this is the point where
dirty black holes behave qualitatively differently from the Kerr metric in
that the dependence of the ISCO radius on $b$ is much stronger than in the
Kerr case.

It is instructive to evaluate the relation between $H(0)$ and $H(\infty )$
for vanishing and large magnetic fields that results, according to (\ref{h}%
), in different values of corresponding radii $x_{0}$. The dependence on the
magnetic field is due to the term $\frac{1}{P}\frac{dP}{dx}$ in the
denominator.

\begin{equation}
\frac{H^{3}(0)}{H^{3}(\infty )}=\frac{2\frac{a_{2}}{a_{1}}D+C+Dw_{b=\infty }%
}{2\frac{a_{2}}{a_{1}}D+C+Dw_{b=0}}\text{, }w\equiv \frac{1}{P}\frac{dP}{dx}%
\text{.}
\end{equation}

One can find that%
\begin{equation}
w_{b=0}=-\frac{2R_{+}^{\prime }}{R_{+}}\frac{D}{d}\text{,}
\end{equation}%
\begin{equation}
w_{b=\infty }=-2\{\frac{R_{+}^{\prime }}{R_{+}}[\frac{2\sqrt{D}+\sqrt{d}}{%
\sqrt{d}+\sqrt{D}}]+\frac{B_{+}^{\prime }}{B_{+}}\}\frac{(\sqrt{d}+\sqrt{D})%
}{\sqrt{d}}.
\end{equation}

Thus for $d\sim D\sim C$, $H(0)\sim H(\infty )$. However, in general they
can differ significantly. Say, for $C=0=B_{+}^{\prime }$ and $d\ll D$, $d\ll
D\frac{R_{+}^{\prime }}{R_{+}}\frac{a_{1}}{a_{2}}$, we have $\frac{H^{3}(0)}{%
H^{3}(\infty )}$ $\approx \frac{2d}{D}\ll 1$. As a result, the ISCO radius (%
\ref{h}) also may vary over wide range.

\section{Slowly rotating black hole}

Now, we assume that $\kappa $ is not small, so the first term in (\ref{N})
dominates. Here, we will consider different cases separately.

\subsection{Non-rotating black hole}

Here, we generalize the results known for the Schwarzschild black hole \cite%
{ag}, \cite{fs}, to a more general metric of a dirty static black hole. In
eqs. (\ref{fd}), (\ref{sd}) we should put $a_{1}=0=a_{2}\,.$ For a finite
value of the magnetic field parameter $b$, ISCO lies at some finite distance
from the horizon. However, now we will show that in the limit $b\rightarrow
\infty ,$ the radius of ISCO tends to that of the horizon with $x_{0}\sim
b^{-1}.$

We will show that this indeed happens, provided the term with $L$ in (\ref{b}%
) is large and compensates the second one with $b$. Correspondingly, we write

\begin{equation}
\mathcal{L}=\mathcal{L}_{0}+\mathcal{L}_{1}\text{,}  \label{l01}
\end{equation}%
where%
\begin{equation}
\frac{\mathcal{L}_{0}}{R_{+}}=b\text{.}  \label{lb}
\end{equation}%
For what follows, we introduce the quantity 
\begin{equation}
\alpha =\frac{\mathcal{L}_{1}}{R_{+}}\text{,}  \label{al}
\end{equation}%
$\alpha =O(1)$. Then, near the horizon, where $x$ is small, we can use the
Taylor expansion

\begin{equation}
\beta =\alpha -2\frac{\beta _{0}}{R_{+}}x-x\alpha \frac{R_{+}^{\prime }}{%
R_{+}}+\frac{\beta _{2}}{R_{+}^{2}}bx^{2}+...\text{,}  \label{expb}
\end{equation}%
\begin{equation}
\beta _{2}=R_{+}^{\prime 2}-R_{+}R_{+}^{\prime \prime }-\frac{R_{+}^{\prime
}R_{+}B_{+}^{\prime }}{B_{+}}-\frac{R_{+}^{2}B_{+}^{\prime \prime }}{2B_{+}}%
\text{,}  \label{b2}
\end{equation}%
where%
\begin{equation}
\beta _{0}=bs\text{, }  \label{bbs}
\end{equation}%
\begin{equation}
s=R_{+}^{\prime }+\frac{1}{2}\frac{B_{+}^{\prime }R_{+}}{B_{+}}\text{.}
\label{sb}
\end{equation}%
Now, $\beta _{0}\gg 1$ but, by assumption, $\beta $ is finite.

In terms of the variable%
\begin{equation}
u=\frac{\beta _{0}}{R_{+}}x\text{,}  \label{ux}
\end{equation}%
it can be rewritten as 
\begin{equation}
\beta =\alpha -2u+\frac{u}{\beta _{0}}(\frac{\beta _{2}u}{s}-\alpha
c)+O(\beta _{0}^{-2})\text{,}  \label{bu}
\end{equation}%
\begin{equation}
c=R_{+}^{\prime }\text{.}  \label{cr}
\end{equation}

It is clear from the above formulas that the expansion with respect to the
coordinate $x$ is equivalent to the expansion with respect to inverse powers
of the magnetic field $b^{-1}$, so for $b\gg 1$ this procedure is reasonable.

Then, after substitution of (\ref{bu}), we can represent (\ref{fd}) and (\ref%
{sd}) in the form of expansion with respect to $\beta _{0}^{-1}$:

\begin{equation}
\frac{1}{m^{2}}\frac{dV_{eff}}{d\rho }=C_{0}+\frac{C_{1}}{\beta _{0}}%
+O(\beta _{0}^{-2})=0,  \label{ff}
\end{equation}%
\begin{equation}
-\frac{1}{2m^{2}}\frac{d^{2}V_{eff}}{d\rho ^{2}}=-S_{1}\beta
_{0}-S_{0}+O(\beta _{0}^{-1})=0\text{.}  \label{ss}
\end{equation}%
Here, the coefficients at leading powers are equal to

\begin{equation}
C_{0}=2\kappa (12u^{2}-8u\alpha +1+\alpha )\text{,}
\end{equation}%
\begin{equation}
S_{1}=16\kappa (3u-\alpha )\text{.}
\end{equation}%
Then, in the main approximation we have equations $C_{0}=0$ and $S_{1}=0$
which give us 
\begin{equation}
u=\frac{1}{\sqrt{3}},\alpha =\sqrt{3},\beta =\frac{1}{\sqrt{3}}.
\label{main}
\end{equation}%
To find the corrections $O(b^{-1})$, we solve eqs. (\ref{ff}) and (\ref{ss})
perturbatively. In doing so, it is sufficient to substitute these values
into further coefficients $C_{1}$ and $S_{0}\,$. The results are listed in
Appendix.

In the particular case of the Schwarschild metric, $c=\beta _{2}=s=1$, $%
D=-r_{+}^{-2}$, $\kappa =(2R_{+})^{-1}$, $R_{+}=r_{+}$. Writing $r_{+}=2M\,$%
, where $M$ is the black hole mass, we have from (\ref{l01}), (\ref{al}), (%
\ref{ux}), (\ref{b1}), (\ref{lb3}) and (\ref{en})%
\begin{equation}
\frac{r_{0}-r_{+}}{M}\approx \frac{2}{\sqrt{3}b}-\frac{8}{3b^{2}}\text{,}
\label{r1}
\end{equation}%
\begin{equation}
\frac{\mathcal{L}}{R_{+}}\approx b+\sqrt{3}-\frac{1}{3b}\text{,}  \label{lsw}
\end{equation}%
\begin{equation}
\mathcal{E}_{0}\approx \frac{2}{3^{3/4}\sqrt{b}}.  \label{ens}
\end{equation}

Eqs.(\ref{r1}), (\ref{ens}) agree with \cite{fr} and \cite{weak}.

It is interesting that in terms of variables $u,$ $\frac{\mathcal{L}_{0}}{%
R_{+}}$ and $b$ the result (\ref{main}) looks model-independent in the main
approximation. This can be thought of as manifestation of the universality
of black hole physics near the horizon. Dependence on a model reveals itself
in higher-order corrections.

\subsection{Extremely slow rotation}

Now, we consider rotation as perturbation. Here, the angular velocity of
rotation is the most small parameter. Correspondingly, in the expressions (%
\ref{2d}), (\ref{sd}) we neglect the term $L^{2}$ since it contains $\omega
^{\prime 2}.$ More precisely, we assume%
\begin{equation}
\mathcal{L}a_{1}^{2}\ll a_{2}N\text{,}  \label{ln}
\end{equation}%
so from (\ref{n0}), (\ref{lb3}) we have%
\begin{equation}
R_{+}b^{3/2}a_{1}^{2}\ll a_{2}\sqrt{\kappa R_{+}}\text{.}  \label{lan}
\end{equation}

In the particular case of the slow rotating Kerr metric, $\kappa \approx 
\frac{1}{2R_{+}}$, $a_{1}\sim \frac{a}{M^{3}}=\frac{a^{\ast }}{M^{2}}$, $%
a_{2}\sim \frac{a}{M^{4}}=\frac{a^{\ast }}{M^{3}}$,where $a=J/M$, $J$ is the
angular momentum of a black hole, $a^{\ast }=\frac{a}{M}$. Then, (\ref{ln})
reads%
\begin{equation}
a^{\ast }b^{3/2}\ll 1\text{.}  \label{abk}
\end{equation}

There are two kinds of corrections - due to the magnetic field and due to
rotation. One can check that the presence of rotation leads to the
appearance in the series (\ref{ff}), (\ref{ss}) of half-integer inverse
powers of $\beta _{0}$, in addition to integer ones. In the main
approximation, we consider both kinds of corrections as additive
contributions. Omitting details, we list the results:%
\begin{equation}
u\approx \frac{1}{\sqrt{3}}+\frac{1}{3bs}(\frac{5}{6}\frac{\beta _{2}}{s}+%
\frac{1}{3}\frac{DR_{+}}{\kappa }-\frac{3c}{2})-\frac{\sqrt{2}\sqrt{R_{+}}}{%
3^{5/4}\sqrt{\kappa }\sqrt{s}}a_{1}R_{+}\sqrt{b}\text{,}
\end{equation}%
\begin{equation}
\frac{\mathcal{L}}{R_{+}}\approx b+\sqrt{3}-\sqrt{2}\frac{1}{3^{3/4}\sqrt{%
\kappa }\sqrt{s}}a_{1}R_{+}^{3/2}\sqrt{b},
\end{equation}%
\begin{equation}
N_{0}\approx \frac{\sqrt{2\kappa R_{+}}}{3^{1/4}\sqrt{bs}}.  \label{n1}
\end{equation}%
It follows from (\ref{X}), (\ref{xx}) that%
\begin{equation}
X\approx \frac{2^{3/2}m}{3^{3/4}}\sqrt{\frac{\kappa R_{+}}{bs}},  \label{xsl}
\end{equation}%
\begin{equation}
\mathcal{E}\approx R_{+}\omega _{+}b+\frac{\sqrt{\kappa R_{+}}}{\sqrt{b}%
\sqrt{s}}\frac{2^{3/2}}{3^{3/4}}\,\text{.}  \label{esr}
\end{equation}

For the slow rotating Kerr metric, $R_{+}\approx 2M$,

\begin{equation}
\omega =\frac{R_{+}a}{r^{3}}+O(a^{2})\text{.}  \label{kom}
\end{equation}%
In \ the main approximation the difference between the Boyer-Lindquist
coordinate $r$ and quasiglobal one $\rho $ has the same order $a^{2}$ and
can be neglected. Then,%
\begin{equation}
a_{1}=\frac{3a^{\ast }}{2R_{+}^{2}}\text{,}  \label{a1}
\end{equation}%
\begin{equation}
u\approx \frac{1}{\sqrt{3}}+\frac{1}{\sqrt{3}b}-\frac{4}{3b^{2}}-\frac{1}{%
3^{1/4}}\frac{a^{\ast }}{\sqrt{b}}\text{,}  \label{ua}
\end{equation}

\begin{equation}
\mathcal{E}\approx \frac{1}{\sqrt{b}}\frac{2}{3^{3/4}}+\frac{a^{\ast }}{2}b%
\text{,}  \label{eab}
\end{equation}%
\begin{equation}
\frac{\mathcal{L}}{R_{+}}\approx b+\sqrt{3}-3^{3/4}a^{\ast }\sqrt{b}.
\label{la}
\end{equation}

They agree with the results of Sec. 3 B 2 of \cite{weak}. It is seen from (%
\ref{ua}) - (\ref{la}) that the fractional corrections have the order $%
a^{\ast }b^{3/2}$ and are small in accordance with (\ref{abk}). In a more
general case, the small parameter of expansion corresponds to (\ref{lan}),
so it is the quantity $\frac{\sqrt{R_{+}}b^{3/2}a_{1}^{2}}{a_{2}\sqrt{\kappa 
}}$.

\subsection{Modestly slow rotation}

Let now, instead of (\ref{ln}), (\ref{lan}) the opposite inequalities hold:

\begin{equation}
\mathcal{L}a_{1}^{2}\gg a_{2}N\text{,}
\end{equation}%
\begin{equation}
R_{+}b^{3/2}a_{1}^{2}\gg a_{2}\sqrt{\kappa R_{+}},
\end{equation}%
or%
\begin{equation}
a^{\ast }b^{3/2}\gg 1  \label{ab3}
\end{equation}%
in the Kerr case. Correspondingly, in what follows the small parameter of
expansion is $\frac{a_{2}\sqrt{\kappa },}{\sqrt{R_{+}}b^{3/2}a_{1}^{2}}$
that reduces to ($a^{\ast }b^{3/2}$)$^{-1}$ in the Kerr case.

Additionally, we assume that%
\begin{equation}
ba_{1}^{\ast 2}\gg 1\text{.}  \label{ba1}
\end{equation}

It turns out (see the details in Appendix) that

\begin{equation}
x_{0}\approx \frac{R_{+}\delta ^{2}}{36}a_{1}^{\ast 2}\text{,}
\end{equation}%
\begin{equation}
\omega _{0}\approx \omega _{+}-\frac{\delta ^{2}}{36R_{+}}a_{1}^{\ast 3}%
\text{,}  \label{oma}
\end{equation}%
\begin{equation}
\frac{\mathcal{L}}{R_{+}}=b(1-\frac{\delta ^{2}}{6}sa_{1}^{\ast 2})\text{,}
\label{lm}
\end{equation}%
\begin{equation}
N_{0}\approx \frac{1}{3\sqrt{2}}\sqrt{\kappa R_{+}}a_{1}^{\ast }\delta \text{%
,}  \label{n2}
\end{equation}%
where $\delta =\frac{1}{s\sqrt{2\kappa R_{+}}}$.

It follows from (\ref{49}), (\ref{hy}), (\ref{hh}) that%
\begin{equation}
\beta _{+}=\beta (0)\approx -\frac{1}{2}\beta _{0}a_{1}^{\ast 2},
\label{be0}
\end{equation}%
\begin{equation}
\beta (x_{0})\approx -2\frac{\delta ^{2}}{9}bsa_{1}^{\ast 2}\text{.}
\label{bh1}
\end{equation}%
\begin{equation}
X_{0}\approx m\frac{1}{27}\sqrt{2\kappa R_{+}}\delta ^{3}bsa_{1}^{\ast 3}%
\text{,}  \label{x27}
\end{equation}%
\begin{equation}
\mathcal{E}_{0}\approx \omega _{+}R_{+}b+\nu ba_{1}^{\ast 3}\text{,}
\label{ea}
\end{equation}%
\begin{equation}
\nu =\frac{1}{27}\sqrt{2\kappa R_{+}}s\delta ^{3}-R_{+}s\delta ^{2}\frac{%
\omega _{+}}{a_{1}^{\ast }}-\frac{\delta ^{2}}{36}\text{.}  \label{nu}
\end{equation}

\subsubsection{Kerr metric}

In the case of the slow rotating Kerr black hole, eq. (\ref{a1}) entails%
\begin{equation}
a_{1}^{\ast }=\frac{3}{2}a^{\ast }\text{,}  \label{3a}
\end{equation}%
$\delta =1=s$,

\begin{equation}
x_{0}\approx \frac{R_{+}}{36}a_{1}^{2\ast }=\frac{R_{+}}{16}a^{\ast 2}\text{,%
}  \label{x16}
\end{equation}%
where we used (\ref{ea}).

One should compare this result to that in \cite{weak}. Now, $R_{+}=2M$, the
horizon radius of the Kerr metric $r_{+}\approx 2M(1-\frac{a^{\ast 2}}{4})$.
Eq. (53) of \cite{weak} gives us%
\begin{equation}
r_{0}\approx 2M(1-\frac{3a^{\ast 2}}{16})\text{,}
\end{equation}%
whence $x_{0}=r_{0}-r_{+}\approx \frac{R_{+}}{16}a^{\ast 2}$ that coincides
with (\ref{x16}). It is seen from (\ref{lm}), (\ref{3a}) that the angular
momentum takes the value%
\begin{equation}
\frac{\mathcal{L}}{R_{+}}\approx b(1-\frac{3}{8}a^{\ast 2})
\end{equation}%
that coincides with eq. (55) of \cite{weak}. Also, one finds that

\begin{equation}
X_{0}\approx \frac{m}{8}ba^{\ast 3}\text{.}
\end{equation}

In eq. (\ref{oma}) one should take into account that $\omega _{+}$ depends
on $r_{+}$ that itself can be expressed in terms of $a^{\ast }$ and $M$.
Collecting all terms, one obtains from (\ref{ea})%
\begin{equation}
\mathcal{E}_{0}\approx \frac{a^{\ast }b}{2}-\frac{ba^{\ast 3}}{32}
\end{equation}%
that agrees with eq. (54) of \cite{weak}.

\section{ISCO for rotating nonextremal black holes in the strong magnetic
field}

In the previous section we saw that in the limit $b\rightarrow \infty $ the
ISCO radius does not coincide with that of the horizon that generalizes the
corresponding observation made in Sec. III B 3 of \cite{weak}. Now, we will
see that this is a general result which is valid for an arbitrary degree of
rotation and finite $\kappa $ (so, for generic nonextremal black holes). It
is worth noting that for $b=0$ it was noticed that the near-horizon ISCO are
absent \cite{piat}, \cite{circ}. However, for $b\gg 1$ the corresponding
reasonings do not apply, so we must consider this issue anew.

We have to analyze eqs. (\ref{1d}), (\ref{2d}) in which (\ref{xx}) is taken
into account.

Neglecting higher order corrections, we can rewrite them in the form

\begin{equation}
(2\kappa +2Dx)(1+\beta ^{2})+(2\kappa x+Dx^{2})\frac{d\beta ^{2}}{dx}-\frac{%
2L\sqrt{1+\beta ^{2}}}{m}N(a_{1}-2a_{2}x)=0\text{,}  \label{d1}
\end{equation}

\begin{equation}
\frac{2L^{2}}{m^{2}}(a_{1}-2a_{2}x)^{2}-2a_{2}N\sqrt{1+\beta ^{2}}\frac{L}{m}%
-W=0\text{,}  \label{d2}
\end{equation}%
\begin{equation}
W=(2D+6Cx)(1+\beta ^{2})+2\frac{d\beta ^{2}}{dx}(2\kappa +2Dx)+(2\kappa
x+Dx^{2})\frac{d^{2}\beta ^{2}}{dx^{2}}\text{.}  \label{w}
\end{equation}

1) Let us suppose that $\beta $ is finite or, at least, $\beta \ll b$. Then,
it follows from (\ref{expb}), (\ref{b}), (\ref{lb}) that $\frac{d\beta }{dx}%
\sim b$ and $L\sim b$. Also, $x\sim b^{-1}$ according to (\ref{ux}), $N\sim 
\sqrt{x}\sim b^{-1/2}$. However, it is impossible to compensate the term
with $L^{2}$ in (\ref{d2}) having the order $b^{2}$.

2) Let $\beta \sim L\sim b$. Then, in (\ref{d1}) the first term has the
order $b^{2}\,$and cannot be compensated.

3) $\beta \gg b$. Then, (\ref{b}) gives us that $L\sim \beta $. Again, the
first term in (\ref{d1}) cannot be compensated.

Thus we see that, indeed, in the limit $b\rightarrow \infty $ the assumption
about $x\rightarrow 0$ leads to contradictions, so ISCO radius does not
approach the horizon.

\section{Extremal nonrotating black hole}

Up to now, we considered the case of a nonextremal black hole, so the
surface gravity $\kappa $ was arbitrary or small quantity but it was nonzero
anyway. Let us discuss now the case of the extremal black hole, so $\kappa
=0 $ exactly. We pose the question: is it possible to get the ISCO such that
for $b\rightarrow \infty $ the ISCO radius tends to that of the horizon?
Now, we will see that this is indeed possible for a nonrotating black hole ($%
\omega =0$). We assume that the electric charge that can affect the metric
is negligible. The extremal horizon appears due to properties of matter that
surrounds the horizon that is possible even in the absence of the electric
charge, provided equation of state obeys some special conditions \cite{bron}.

For ISCO close to the horizon we can use the expansion%
\begin{equation}
N^{2}=Dx^{2}+...  \label{nd}
\end{equation}%
in which we drop the terms of the order $x^{3}$ and higher. Now we show that
the case under discussion does exist with a finite quantity $\beta $. We can
use now (\ref{bu}) in which only the first term is retained, so%
\begin{equation}
\beta \approx \alpha -2u\text{,}
\end{equation}%
where $u$ is given by eq. (\ref{ux}). Then, (\ref{v}) reads 
\begin{equation}
V\approx m^{2}\frac{DR_{+}^{2}}{b^{2}s^{2}}f(u)-E^{2}\text{,}
\end{equation}%
where%
\begin{equation}
f(u)=u^{2}(1+\alpha ^{2}-4u\alpha +4u^{2}).
\end{equation}%
Eqs. (\ref{fd}), (\ref{sd}) reduce to

\begin{equation}
\frac{df}{du}(u_{0})=0\text{,}
\end{equation}%
\begin{equation}
\frac{d^{2}f}{du^{2}}(u_{0})=0\text{.}
\end{equation}%
They have the solution%
\begin{equation}
u_{0}=\frac{3}{2^{3/2}}\text{, }\alpha =\frac{4}{\sqrt{2}}=2\sqrt{2}\text{,}
\end{equation}%
whence 
\begin{equation}
\beta \approx \frac{1}{\sqrt{2}}.  \label{eb2}
\end{equation}%
Correspondingly, eqs. (\ref{nd}), eq. (\ref{0}) give us%
\begin{equation}
N(x_{0})\approx \frac{3}{2^{3/2}}\sqrt{D}\frac{R_{+}}{bs},  \label{ne}
\end{equation}%
\begin{equation}
\frac{X_{0}}{m}=\mathcal{E}_{0}\approx \frac{3^{3/2}}{4}\sqrt{D}\frac{R_{+}}{%
bs}\text{.}  \label{ee}
\end{equation}

We can also find the angular momentum on ISCO%
\begin{equation}
\frac{\mathcal{L}_{0}}{R_{+}}\approx b+\frac{1}{2}\sqrt{2}.
\end{equation}

Thus for big $b$ there is ISCO outside the horizon that tends to it in the
limit $b\rightarrow \infty $, so that the quantity $x_{0}\rightarrow 0$.

\section{Extremal rotating black hole}

Now, we consider the same question but now for rotaitng black holes: is it
possible to have ISCO in the near-horizon region (as closely as we like) for
the extremal BH, when $\kappa =0$? Mathematically, it would mean that%
\begin{equation}
\lim_{b\rightarrow \infty }x_{0}=0\text{.}
\end{equation}

Then, (\ref{1d}), (\ref{nx}) with $\kappa =0$ give us for small $x$ that%
\begin{equation}
x_{0}D[(1+\beta ^{2})-\frac{La_{1}\sqrt{1+\beta ^{2}}}{m\sqrt{D}}]+\frac{%
x_{0}^{2}}{2}\{C[3(1+\beta ^{2})-\frac{La_{1}\sqrt{1+\beta ^{2}}}{m\sqrt{D}}%
]+D\left( \beta ^{2}\right) ^{\prime }\}=0.  \label{s1}
\end{equation}%
Eq. (\ref{2d}) with terms of the order $x_{0}^{2}$ and higher neglected,
gives rise to%
\begin{equation}
D(1+\beta ^{2})-\mathcal{L}^{2}a_{1}^{2}+x_{0}[2D\left( \beta ^{2}\right)
^{\prime }+2a_{2}\mathcal{L}^{2}\sqrt{D}\sqrt{1+\beta ^{2}(x_{0})}+2\mathcal{%
L}^{2}a_{1}a_{2}+3C(1+\beta ^{2})]=0.  \label{s}
\end{equation}

Then, the main terms in (\ref{s1}), (\ref{s}) entail%
\begin{equation}
\mathcal{L}a_{1}=\sqrt{D}\sqrt{1+\beta ^{2}}\text{.}  \label{lad}
\end{equation}%
For $b\gg 1$, assuming for definiteness that $d>D$ ($d$ is defined according
to (\ref{da})), one finds from (\ref{b}) and (\ref{lad}) that

\begin{equation}
\beta _{+}\approx -b\frac{\sqrt{d}}{\sqrt{d}+\sqrt{D}}\text{,}
\end{equation}%
\begin{equation}
\frac{\mathcal{L}}{R_{+}}=b\frac{\sqrt{D}}{\sqrt{d}+\sqrt{D}}\text{.}
\end{equation}%
\begin{equation}
\frac{L^{2}}{m^{2}}a_{1}^{2}=D+D(\frac{L^{2}}{m^{2}R_{+}^{2}}-2\frac{L}{%
mR_{+}}b+b^{2})
\end{equation}

The terms $x_{0}^{2}$ in (\ref{s1}) and $x_{0}$ in (\ref{s}) give us, with (%
\ref{lad}) taken into account%
\begin{equation}
(1+\beta ^{2})C+D\left( \beta ^{2}\right) ^{\prime }=0,
\end{equation}

\begin{equation}
\left( \beta ^{2}\right) ^{\prime }+2\frac{a_{2}}{a_{1}}(1+\beta ^{2})+\frac{%
3C}{2D}(1+\beta ^{2})=0\text{,}
\end{equation}%
whence%
\begin{equation}
C=-4D\frac{a_{2}}{a_{1}}\text{.}  \label{cd}
\end{equation}

The system is overdetermined, eq. (\ref{cd}) cannot be satisfied in general.
In principle, one can consider (\ref{cd}) as restriction on the black hole
parameters. This is similar to the situation for the extremal Kerr-Newman
metric ($b=0$), where ISCO near the horizon exists only for the selected
value of the angular momentum approximately equal to $\frac{a}{M}\approx 
\frac{1}{\sqrt{2}}$ \cite{m}, \cite{extc}. However, we will not discuss such
exceptional cases further. Generically, the answer to our question is
negative, so the ISCO radius does not approach the horizon in the limit $%
b\rightarrow \infty .$

\section{Particle collisions: general formulas}

Let two particles collide. We label their characteristics by indices 1 and
2. Then, in the point of collision, one can define the energy in the centre
of mass (CM) frame as%
\begin{equation}
E_{c.m.}^{2}=-p_{\mu }p^{\mu }=m_{1}^{2}+m_{2}^{2}+2m_{1}m_{2}\gamma \text{.}
\end{equation}

Here,%
\begin{equation}
p^{\mu }=m_{1}u_{1}^{\mu }+m_{2}u_{2}^{\mu }
\end{equation}%
is the total momentum,%
\begin{equation}
\gamma =-u_{1\mu }u_{2}^{\mu }
\end{equation}%
is the Lorentz factor of their relative motion.

For motion in the equatorial plane in the external magnetic field (\ref{ab}%
), one finds from the equations of motion (\ref{ft}), (\ref{rt}) that%
\begin{equation}
\gamma =\frac{X_{1}X_{2}-\varepsilon _{1}\varepsilon _{2}\sqrt{V_{1}V_{2}}}{%
m_{1}m_{2}N^{2}}-\beta _{1}\beta _{2}\text{.}  \label{ga}
\end{equation}

Here, $\varepsilon =+1$, if the particle moves away from the horizon and $%
\varepsilon =-1$, if it moves towards it.

Now, there are two scenarios relevant in our context. We call them
O-scenario and H-scenario according to the terminology of \cite{circ}.
Correspondingly, we will use superscripts "O" and "H".

\subsection{O - scenario}

Particle 1 moves on ISCO. As $V_{1}(\rho _{0})=0$ on ISCO, the formula
simplifies to%
\begin{equation}
(E_{c.m.}^{O})^{2}=m_{1}^{2}+m_{2}^{2}+2(\frac{X_{1}X_{2}}{N^{2}}%
-m_{1}m_{2}\beta _{1}\beta _{2})\text{.}  \label{o}
\end{equation}

As we are interested in the possibility to get $\gamma $ as large as one
likes, we will consider the case when the ISCO is close to the horizon, so $%
N $ is small. In doing so, we will assume that $(X_{2})\neq 0$, so particle
2 is usual according to the terminology of \cite{circ}. We also must take
into account eq. (\ref{xx}), whence%
\begin{equation}
(E_{c.m.}^{O})^{2}=m_{1}^{2}+m_{2}^{2}+2(m_{1}\frac{X_{2}\sqrt{1+\beta
_{1}^{2}}}{N}-m_{1}m_{2}\beta _{1}\beta _{2})\text{.}  \label{gis}
\end{equation}

For ISCO close to the horizon, the first term dominates and we have%
\begin{equation}
(E_{c.m.}^{O})^{2}\approx 2m_{1}\frac{(X_{2})_{0}(\sqrt{1+\beta _{1}^{2}}%
)_{0}}{N_{0}}\text{.}  \label{go}
\end{equation}

\subsection{H - scenario}

Now, particle 1 leaves ISCO (say, due to additional collision) with the
corresponding energy $E=E(x_{0})$ and angular momentum $L=L(x_{0})$ that
corresponds just to ISCO. This particle moves towards the horizon where it
collides with particle 2.

Mathematically, it means that we should take the horizon limit $N\rightarrow
0$ first in formula (\ref{ga}). We assume that both particles move towards
the horizon, so $\varepsilon _{1}\varepsilon _{2}=+1$. Then,%
\begin{equation}
(E_{c.m.}^{H})^{2}=m_{1}^{2}+m_{2}^{2}+m_{1}^{2}(1+\beta _{1}^{2})\frac{X_{2}%
}{X_{1}}+m_{2}^{2}(1+\beta _{2}^{2})\frac{X_{1}}{X_{2}}-2m_{1}m_{2}\beta
_{1}\beta _{2},  \label{hor}
\end{equation}%
where all quantities are to be calculated on the horizon.

For small $X_{1}$, when 
\begin{equation}
X_{1}\ll X_{2}\frac{m_{1}}{m_{2}}\sqrt{\frac{1+\beta _{1}^{2}}{1+\beta
_{2}^{2}}},
\end{equation}

we see from (\ref{hor}) that%
\begin{equation}
(E_{c.m.}^{H})^{2}\approx m_{1}^{2}(1+\beta _{1}^{2})_{+}\frac{(X_{2})_{+}}{%
(X_{1})_{+}}\text{.}  \label{e1}
\end{equation}%
Now, 
\begin{equation}
X_{1}=E_{0}-\omega _{+}L_{0}=X_{0}+(\omega _{0}-\omega _{+})L_{0}\text{,}
\end{equation}%
where $X_{0}=E_{0}-\omega _{0}L$ corresponds to ISCO. With (\ref{om}), (\ref%
{xx}) taken into account, in the main approximation%
\begin{equation}
X_{1}\approx m_{1}N_{0}\sqrt{1+\beta _{1}^{2}(x_{0})}-a_{1}x_{0}L_{0}\text{.}
\label{xis}
\end{equation}

Now, we apply these formulas to different cases considered above.

\section{Kinematics of motion on ISCO}

It is instructive to remind that the general explanation of high $E_{c.m.}$
consists in the simple fact that a rapid usual particle having a velocity
close to speed of light, hits the slow particle that has parameters
approximately equal the critical values. This was explained in detail in 
\cite{k} for the standard BSW effect (without considering collision near
ISCO). Does this explanation retain its validity in the present case? One
particle that participates in collision is usual, so it would cross the
horizon with the velocity approaching the speed of light in an appropriate
stationary frame (see below). We consider the near-horizon ISCO, so the
velocity of a usual particle is close to the speed of light. Now, we must
check what happens to the velocity of a particle on ISCO.

To describe kinematic properties, it is convenient to introduce the tetrads
that in the local tangent space enable us to use formulas similar to those
of special relativity. A natural and simple choice is the tetrad of
so-called zero-angular observer (ZAMO) \cite{72}. It reads%
\begin{equation}
h_{(0)\mu }=-N(1,0,0,0)\text{, }  \label{h0}
\end{equation}%
\begin{equation}
h_{(1)\mu }=N^{-1}(0,1,0,0),
\end{equation}%
\begin{equation}
h_{(2)\mu }=\sqrt{g_{\theta }}(0,0,1,1),
\end{equation}%
\begin{equation}
h_{(3)\mu }=R(-\omega ,0,0,1)\text{.}  \label{h3}
\end{equation}%
Here, $x^{0}=t,x^{1}=r$, $x^{2}=\theta $, $x^{3}=\phi $. It is also
convenient to define the local three-velocity \cite{72} according to%
\begin{equation}
v^{(a)}=v_{(a)}=\frac{u^{\mu }h_{\mu (a)}}{-u^{\mu }h_{\mu (0)}},  \label{vi}
\end{equation}%
$a=1,2,3$.

From equations of motion (\ref{0}) - (\ref{1})\ and formulas for tetrad
components, we obtain

\begin{equation}
-u^{\mu }h_{\mu (0)}=\frac{X}{mN},
\end{equation}%
\begin{equation}
u^{_{\mu }}h_{\mu (3)}=\beta \text{,}
\end{equation}%
\begin{equation}
v^{(3)}=\frac{m\beta N}{X},  \label{v3}
\end{equation}%
\begin{equation}
v^{(1)}=\sqrt{1-\frac{m^{2}N^{2}}{X^{2}}(1+\beta ^{2})},  \label{v1}
\end{equation}%
the component $v^{(2)}=0$ for equatorial motion.

Then, introducing also the absolute value of the velocity $v$ according to%
\begin{equation}
v^{2}=\left[ v^{(1)}\right] ^{2}+\left[ v^{(3)}\right] ^{2}
\end{equation}%
one can find that%
\begin{equation}
X=m\gamma _{0}N\text{, }\gamma _{0}=\frac{1}{\sqrt{1-v^{2}}}\text{.}
\label{xv}
\end{equation}

Eq. (\ref{xv}) was derived in \cite{k} for the case when the magnetic field
is absent. We see that its general form does not depend on the presence of
such a field.

For a circle orbit, eq.(\ref{xx}) should hold. Comparing it with (\ref{xv}),
we find that%
\begin{equation}
\gamma _{0}=\sqrt{1+\beta ^{2}}
\end{equation}%
that has the same form as for the static case \cite{fr}. Now we can consider
different cases depending on the value of the magnetic field and a kind of a
black hole.

\subsection{Near-extremal black holes}

For small $b$, eq. (\ref{bsm}) shows that $\beta $ is finite, so is the
quantity $\gamma _{0}$. Therefore, $v<1$. For the Kerr metric, in the main
approximation, $\beta =\frac{L_{0}}{mR_{+}}=\frac{1}{\sqrt{3}}$ on ISCO (see
eqs. \ 4.7 of \cite{circkerr} and eq. 77 of \cite{circ}, $R_{+}=2M$), so $%
\gamma _{0}=\frac{2}{\sqrt{3}}$, $v=\frac{1}{2}$ that is a known result (see
discussion after eq. 3.12 b in \cite{72}).

For \ large $b$, accoding to (\ref{bbs}), the quantity $\beta \,\ $is
proportional to $b,$and grows, $v\rightarrow 1$. However, this case is not
very interesting since the individual energy (\ref{el}) diverges itself.

\subsection{Non-rotating or slowly rotating nonextremal black \ holes}

According to eq. (\ref{main}), $\beta \approx \frac{1}{\sqrt{3}}$. Slow
rotation adds only small corrections to this value. Thus, rather
unexpectedly, we again obtain that on ISCO%
\begin{equation}
v\approx \frac{1}{2}.  \label{12}
\end{equation}%
This value coincides for the near-extremal Kerr without a magnetic field and
a nonrotating or slow rotating dirty black hole in the strong magnetic field.

\subsection{Modestly rotating nonextremal black hole}

It follows from (\ref{be0}), (\ref{ba1}) that $\left\vert \beta \right\vert
\gg 1$. However, as the energy of a particle on ISCO (\ref{ea}) tends to
inifnity, this case is also not so interesting.

To summarize, in all cases of interest (when an individual energy is
finite), $\beta $ remains finite even in the strong magnetic field.
Correspondingly, $v<1$ on ISCO and the previous explanation of the high $%
E_{c.m.}$ \cite{k} applies. For less interesting cases, when an individual
energy diverges, we have collision between two rapid particles but their
velocities are not parallel and this also gives rise to high $\gamma _{0}$
(see eq. 20 in \cite{k}).

\section{Centre-of mass energy of collision}

\subsection{Near-extremal black hole}

\subsubsection{O - scenario}

Using (\ref{go}), (\ref{nx}) and (\ref{nk}) one obtains%
\begin{equation}
(E_{c.m.}^{O})^{2}\approx 2m_{1}\frac{(X_{2})_{+}\sqrt{1+\beta _{1}^{2}}}{%
\sqrt{D}H\kappa ^{2/3}}\text{.}
\end{equation}

In the strong magnetic field, with $b\gg 1$, using the expression (\ref{bs})
for $\beta $, we obtain 
\begin{equation}
(E_{c.m.}^{O})^{2}\approx \frac{2m_{1}(X_{2})_{+}}{\sqrt{D}H\kappa ^{2/3}}%
\frac{R_{+}a_{1}b}{R_{+}a_{1}+\sqrt{D}}\text{.}
\end{equation}

In the near-extremal Kerr case, $D=M^{-2}$, $R_{+}=2M$, $\kappa \approx 
\frac{1}{2}\sqrt{1-a^{\ast 2}},$ $H=M^{-5/3}$, $a_{1}=M^{-2}$. As \ a result,%
\begin{equation}
(E_{c.m.}^{O})^{2}\approx \frac{2^{8/3}m_{1}(X_{2})_{+}b}{3(1-a^{\ast
2})^{1/3}}
\end{equation}%
that coincides with eq. (61) of \cite{weak} in which the limit $b\rightarrow
\infty $ should be taken.

\subsubsection{H - scenario}

Now, due to (\ref{lad}), eq. (\ref{xis}) gives us $X_{1}=0$. It means that
in the expansion (\ref{nx}) we must retain the first correction in the
expression for $N$, when it is substituted into (\ref{xis}). As a result, we
have%
\begin{equation}
X_{1}\approx \frac{m_{1}\kappa \sqrt{1+\beta _{1}^{2}(x_{0})}}{\sqrt{D}}%
\text{.}
\end{equation}%
There are also terms of the order $x_{0}^{2}\sim \kappa ^{4/3}$ but they are
negligible as compared to $\kappa $. Correspondingly, (\ref{hor}) gives us

\begin{equation}
(E_{c.m.}^{H})^{2}\approx m_{1}\sqrt{D}\sqrt{1+\beta _{1}^{2}}\left(
X_{2}\right) _{+}\kappa ^{-1}\text{.}
\end{equation}

In the strong magnetic field, with $b\gg 1$, using (\ref{bs}) again we obtain

\begin{equation}
(E_{c.m.}^{H})^{2}\approx m_{1}\sqrt{D}\frac{R_{+}a_{1}b}{R_{+}a_{1}+\sqrt{D}%
}\left( X_{2}\right) _{+}\kappa ^{-1}\text{.}
\end{equation}

Thus in both versions, for $b\gg 1$ the effect is enhanced due to the factor 
$b$. For $b=0$ we return to \cite{circ}.

In the Kerr case,

\begin{equation}
(E_{c.m.}^{H})^{2}\approx \frac{4}{3}m_{1}\frac{\left( X_{2}\right) _{+}b}{%
(1-a^{\ast 2})^{1/2}}
\end{equation}%
that corresponds to eq. (59) of \cite{weak} in which $b\gg 1$.

\subsection{Extremely slow rotating or nonrotating black hole}

\subsubsection{O - scenario}

Now, (\ref{n0}) and (\ref{go}) give us%
\begin{equation}
(E_{c.m.}^{O})^{2}\approx 3^{-1/4}\frac{4m_{1}(X_{2})_{+}}{\sqrt{2\kappa
R_{+}}}\sqrt{bs}.  \label{os}
\end{equation}

In the Schwarschild case, $2\kappa R_{+}=1=s$,%
\begin{equation}
(E_{c.m.}^{O})^{2}\approx \frac{4m_{1}(X_{2})_{+}}{3^{1/4}}\sqrt{b}
\end{equation}%
that coincides with eq. (63) of \cite{weak}.

\subsubsection{H - scenario}

Using (\ref{xsl}) and neglecting in (\ref{xis}) the second term (rotational
part), we get%
\begin{equation}
(E_{c.m.}^{H})^{2}\approx \frac{2}{3^{3/4}}\frac{m_{1}(X_{2})_{+}}{\sqrt{%
2\kappa R_{+}}}\sqrt{bs}\text{.}  \label{hs}
\end{equation}

\subsection{Modestly rotating black holes in strong magnetic field}

\subsubsection{O - scenario}

With $\beta _{1}\gg 1$, it follows from (\ref{n2}), (\ref{bh1}) and (\ref{go}%
) that

\begin{equation}
(E_{c.m.}^{O})^{2}\approx \frac{4}{3}\sqrt{2}\frac{m_{1}}{\sqrt{\kappa R_{+}}%
}\delta bs(X_{2})_{+}a_{1}^{\ast }\text{.}
\end{equation}

In the Kerr case, taking into account (\ref{3a}), we obtain

\begin{equation}
(E_{c.m.}^{O})^{2}\approx 4m_{1}(X_{2})_{+}a^{\ast }b
\end{equation}%
that agrees with eq. (67) of \cite{weak}.

\subsubsection{H - scenario}

In a similar manner, one can obtain from (\ref{e1}), (\ref{xis}) and (\ref%
{x27}) that $(E_{c.m.}^{H})^{2}\sim b$ with a somewhat cumbersome
coefficient that we omit here.

Both these scenarios are less interesting since according to (\ref{ea}), the
individual energy $\mathcal{E}_{0}\sim b$ diverges itself in the limit $%
b\rightarrow \infty $.

\subsection{Extremal nonrotating black holes}

\subsubsection{O - scenario}

Using (\ref{ne}), (\ref{eb2}) we find from (\ref{go}) that%
\begin{equation}
(E_{c.m.}^{O})^{2}\approx \frac{4m_{1}(X_{2})_{+}bs}{\sqrt{3D}R_{+}}\text{.}
\end{equation}

\subsubsection{H - scenario}

Now, it follows from (\ref{ee}), (\ref{eb2}) and (\ref{e1}) that%
\begin{equation}
(E_{c.m.}^{H})^{2}\approx 2m_{1}\frac{(X_{2})_{+}bs}{\sqrt{3D}R_{+}}\text{.}
\end{equation}%
$\frac{b(1+\xi ^{2})}{\Lambda _{+}^{3}}$

\section{Backreaction of magnetic field: Ernst static black hole}

Now, we illustrate the obtained results using the metric of a static
magnetized black hole \cite{ernst} that can be considered as the
generalization of the Schwarzschild solution. This will also allow us to
elucidate the role of backreaciton due to the magnetic field on the behavior
of $E_{c.m.}$ that bounds the BSW effect. The metric reads%
\begin{equation}
ds^{2}=\Lambda ^{2}[-(1-\frac{r_{+}}{r})dt^{2}+\frac{dr^{2}}{1-\frac{r_{\_}}{%
r}}+r^{2}d\theta ^{2}]+\frac{r^{2}\sin ^{2}\theta }{\Lambda ^{2}}d\phi ^{2}%
\text{, }\Lambda ^{2}=1+\frac{B^{2}r^{2}}{4}\sin ^{2}\theta  \label{me}
\end{equation}%
\begin{equation}
A^{\phi }=\frac{\tilde{B}}{2}\text{, }\tilde{B}=B\Lambda \text{,}
\end{equation}%
$r_{+}=2M$ is the horizon radius, $B$ is a constant parameter. It follows
from (\ref{bb}) (with $B$ replaced with $\tilde{B}$) and (\ref{me}) that%
\begin{equation}
\beta =\frac{\mathcal{L}}{R}-b\frac{r}{2M}\text{, }b=\frac{qBM}{m}\text{.}
\end{equation}

Many important details of particle's motion in this background can be found
in Ref. \cite{galpet}.

Calculating the corresponding coefficients according to (\ref{b2}) - (\ref%
{sb}) and substituting them into (\ref{u1}) - (\ref{en}), we obtain%
\begin{equation}
\frac{(r-r_{+})}{r_{+}}=\frac{(1+\xi )}{(1+\xi ^{2})b}[\frac{1}{\sqrt{3}}+%
\frac{-8+18\xi -3\xi ^{2}-2\xi ^{3}-\xi ^{4}}{18b(1+\xi ^{2})^{2}}]
\label{re}
\end{equation}%
\begin{equation}
\xi =B^{2}M^{2}\text{, }
\end{equation}%
\begin{equation}
\frac{\mathcal{L}}{2M}=\frac{1}{(1+\xi )}[b+\sqrt{3}+\frac{-1+3\xi -\xi
^{3}-\xi ^{4}}{3b(1+\xi ^{2})^{2}}]+O(b^{-2})  \label{le}
\end{equation}%
\begin{equation}
\mathcal{E}_{0}\approx \frac{2(1+\xi )^{3/2}}{3^{3/4}\sqrt{b}}\frac{1}{\sqrt{%
1+\xi ^{2}}}\text{.}
\end{equation}%
It follows from (\ref{bet}) that%
\begin{equation}
\beta \approx \frac{1}{\sqrt{3}}+\frac{(\xi -1)(-\xi ^{3}-\xi ^{2}-2\xi +2)}{%
3b(1+\xi ^{2})^{2}}
\end{equation}

For the energy of collision we have from (\ref{os}), (\ref{hs})

\begin{equation}
(E_{c.m.}^{O})^{2}\approx 3^{-1/4}4m_{1}(X_{2})_{+}z,  \label{oe}
\end{equation}

\begin{equation}
(E_{c.m.}^{H})^{2}\approx \frac{2}{3^{3/4}}m_{1}(X_{2})_{+}z\text{,}
\label{he}
\end{equation}%
where%
\begin{equation}
z=\sqrt{\frac{b(1+\xi ^{2})}{(1+\xi )^{3}}}\text{.}
\end{equation}

When $\xi \ll 1$, there is agreement with the results for the Schwarzschild
metric \cite{fr}, \cite{weak} since eq. (\ref{re}) turns into (\ref{r1}) and
(\ref{le}) turns into (\ref{lsw}). It is interesting that for any $\xi $,
the velocity of a particle on ISCO is equal to $1/2$ like this happens for $%
\xi \ll 1.$ The approach under discussion works well also for $\xi \gg 1$,
provided that the ISCO lies close to the horizon to ensure large $E_{c.m.}$,
i,e, $N^{2}\ll 1.$According to (\ref{me}), (\ref{re}), this requires%
\begin{equation}
b\gg \xi
\end{equation}%
or, equivalently,%
\begin{equation}
\frac{q}{m}\gg BM\gg 1\text{.}
\end{equation}

Otherwise, both energies (\ref{oe}), (\ref{he}) contain the factor $\sqrt{%
\frac{b}{\xi }}\sim \sqrt{\frac{q}{mBM}}$ that bounds $E_{c.m.}$ which
begins to decrease when $B$ increases. One should also bear in mind that it
is impossible to take the limit $\xi \rightarrow \infty $ literally since
the geometry becomes singular. In particular, the component of the curvature
tensor $R_{\theta \phi }^{\theta \phi }$ grows like $\xi ^{2}$. The maximum
possible $E_{c.m.}$ is achieved when $\xi \sim 1$, then $z\sim \sqrt{\frac{q%
}{m}}$.

The example with the Ernst metric shows that strong backreaction of the
magnetic field on the geometry may restrict the growth of $E_{c.m.}$ to such
extent that even in spite of large $b$, the effect disappears because of the
factor $\xi $ that enters the metric. It is of interest to consider the
exact rotating magnetized black hole \cite{ek} that generalizes the Kerr
metric but this problem certainly needs separate treatment.

\section{Summary and conclusion}

We obtained characteristics of ISCO and the energy in the CM frame in two
different situations. For the near-extremal case, we considered the BSW
effect. Previous results applied to the weakly magnetized Kerr metric or
dirty black holes without the magnetic field. Now, we took into account both
factors, so generalized the previous results for the case when both matter
and magnetic field are present. In doing so, there is qualitative difference
between dirty rotating black holes and the Kerr one. Namely, the radius of
ISCO depends on the magnetic field strength $b$ already in the main
approximation with respect to small surface gravity $\kappa $ in contrast to
the case of the vacuum metric \cite{weak}, where this dependence reveals
itself in the small corrections only.

For extremal black holes, we showed that, due to the strong magnetic field,
there exists the near-horizon ISCO that does not have a counterpart in the
absence of this field. Correspondingly, we described the effect of
high-energy collisions near these ISCO.

We demonstrated that rotation destroys near-horizon ISCO both for the
nonextremal and extremal horizons. so $\lim_{b\rightarrow \infty
}r_{0}(b)\neq r_{+}$. However, if the parameter responsible for rotation is
small, $E_{c.m.}$ is large in this limit.

For slowly rotating black holes we analyzed two different regimes of
rotation thus having generalized previous results on the Kerr metric \cite%
{weak}. The parameters of expansion used in calculations and the results
agree with the Kerr case. In particuar, for modestly slow rotation the
individal energy of the particle on ISCO is unbound.

In the main approximation, the expressions for the ISCO radius and angular
momentum in dimensionless variables are model-independent, so here one can
see universality of black hole physics.

We also found the three-velocity of a particle on ISCO in the ZAMO frame. It
turned out that for slowly rotating dirty black holes in the magnetic field
it coincides with the value typical of the Kerr metric without a magnetic
field, $v\approx \frac{1}{2}$. Correspondingly, previous explanation of the
high $E_{c.m.}$ as the result of collision of very fast and slow particles 
\cite{k} retains its validity in the scenarios under discussion as well.

In previous studies of the BSW effect in the magnetic field \cite{fr}, \cite%
{weak}, \cite{tur2}, some fixed background was chosen. In this sense, the
magnetic field was supposed to be weak in that it did not affect the metric
significantly (although it influenced strongly motion of charged particles).
Meanwhile, the most part of the formulas obtained in the present work
applies to generic background and only asymptotic behavior of the metric
near the horizon was used. Therefore, they apply to the backgrounds in which
the magnetic field enters the metric itself, with reservation that the
surface gravity $\kappa =\kappa (b)$, etc. In particular, we considered the
static magnetized Ernst black hole and showed that strong backreaction of
the magnetic field on the geometry bounds the growth of $E_{c.m.}$.

Thus we embedded previous scenarios of high energy collisions in the
magnetic field near ISCO in the vicinity of black holes \cite{fr}, \cite%
{weak} and took into account the influence of the magnetic field on the
metric.

Throughout the paper, it was assumed that the effect of the electric charge
on the metric is negligible. It is of interest to extend the approach of the
present work to the case of charged black holes.

\begin{acknowledgments}
This work was funded by the subsidy allocated to Kazan Federal University
for the state assignment in the sphere of scienti c activities.
\end{acknowledgments}

\section{Appendix}

Here, we list some rather cumbersome formulas which are excluded from the
main text.

\subsection{Non-rotating black holes}

\begin{equation}
C_{1}=4\kappa \frac{\beta _{2}}{3\sqrt{3}s}+\frac{4}{3\sqrt{3}}DR_{+},
\end{equation}%
\begin{equation}
S_{0}\approx 8\frac{\kappa }{R_{+}}(-\frac{\beta _{2}}{s}+3c)\text{.}
\end{equation}

The results with the leading term and subleading corrections read%
\begin{equation}
u\approx \frac{1}{\sqrt{3}}+\varepsilon _{1}\text{, }\varepsilon _{1}=\frac{1%
}{3bs}(\frac{5}{6}\frac{\beta _{2}}{s}+\frac{1}{3}\frac{DR_{+}}{\kappa }-%
\frac{3c}{2})\text{,}  \label{u1}
\end{equation}%
\begin{equation}
\alpha \approx \sqrt{3}+\delta _{1}\text{, }\delta _{1}=\frac{1}{3bs}(\frac{%
\beta _{2}}{s}+\frac{DR_{+}}{\kappa })\text{,}  \label{b1}
\end{equation}

\begin{equation}
N\approx \sqrt{2\kappa x_{0}}\approx \sqrt{\frac{2\kappa R_{+}}{bs}}\frac{1}{%
3^{1/4}}\text{,}  \label{n0}
\end{equation}%
\begin{equation}
\frac{\mathcal{L}}{R_{+}}\approx b+\sqrt{3}+\delta _{1}\text{,}  \label{lb3}
\end{equation}%
\begin{equation}
\beta =\frac{1}{\sqrt{3}}+\frac{1}{bs}(\frac{\beta _{2}}{s}\frac{1}{3}%
-c)+O(b^{-2})\text{,}  \label{bet}
\end{equation}%
\begin{equation}
\mathcal{E}_{0}=X_{0}\approx \frac{2^{3/2}}{3^{3/4}}\sqrt{\frac{\kappa R_{+}%
}{bs}}.  \label{en}
\end{equation}

\subsection{Modestly slow rotation}

Now, one can check that, in contrast to the previous case, a finite $\beta $
is inconsistent with eqs. (\ref{ff}), (\ref{ss}). Instead, $\beta \sim b$
for large $b$. By trial and error approach, one can find that the suitable
ansatz reads

\begin{equation}
x=\frac{4}{9}R_{+}ya_{1}^{\ast 2}\text{, }  \label{xe}
\end{equation}%
where we introduced in this ansatz the dimensionless quantity 
\begin{equation}
a_{1}^{\ast }=R_{+}^{2}a_{1},
\end{equation}%
and the coefficient $\frac{4}{9}$ to facilitate comparison to the case of
the Kerr metric (otherwise, this coefficient can be absorbed by $y$).

In doing so,%
\begin{equation}
\beta =\frac{4}{9}\beta _{0}a_{1}^{\ast 2}h(y)\text{, }  \label{49}
\end{equation}%
\begin{equation}
h\approx h_{1}-2y  \label{hy}
\end{equation}%
that is analogue of (\ref{bu}), $\beta _{0}=bs$ according to (\ref{bbs}). By
definition, here $h\neq 0$. For the angular momentum we have from (\ref{b})%
\begin{equation}
\frac{\mathcal{L}}{R_{+}}=b+\beta _{+}=b+\frac{4}{9}\beta _{0}a_{1}^{\ast
2}h_{1}\text{.}  \label{hl}
\end{equation}

Let us consider the main approximation with respect to the parameter $%
\varepsilon =\frac{4}{9}a_{1}^{\ast 2}$ and take into account that $%
\left\vert \beta \right\vert \gg 1$. Then, eq. (\ref{1d}) gives us

\begin{equation}
\left\vert \beta \right\vert \mathcal{L}\sqrt{2\kappa x}a_{1}\approx \kappa
\beta ^{2}+\kappa x\frac{d\beta ^{2}}{dx}.  \label{l1}
\end{equation}

Eq. (\ref{2d}) reads%
\begin{equation}
\mathcal{L}^{2}a_{1}^{2}\approx 2\frac{d\beta ^{2}}{dx}\kappa +\kappa x\frac{%
d^{2}\beta ^{2}}{dx^{2}}\text{,}  \label{l2}
\end{equation}%
where now, with a given accuracy,%
\begin{equation}
\frac{d\beta ^{2}}{dx}=\frac{4}{R_{+}}\beta _{0}^{2}\varepsilon (2y-h_{1}),
\end{equation}%
\begin{equation}
\frac{d\beta ^{2}}{dx^{2}}=\frac{\beta _{0}^{2}}{R_{+}^{2}}\frac{dh^{2}}{%
dy^{2}}=\frac{8\beta _{0}^{2}}{R_{+}^{2}}\text{.}
\end{equation}%
Substituting $\frac{\mathcal{L}}{R_{+}}\approx b$ into (\ref{l1}), (\ref{l2}%
) and assuming $h<0$, one finds the system of two equations 
\begin{equation}
6y-h_{1}=3\delta \sqrt{y}\text{,}  \label{root}
\end{equation}

\begin{equation}
h_{1}=3y-\frac{9}{16}\delta ^{2},  \label{h1}
\end{equation}

$\delta =\frac{1}{s\sqrt{2\kappa R_{+}}}$.This system can be solved easily.
There are two roots here but only one of them satisfies the condition $h\neq
0$:%
\begin{equation}
y_{0}=\frac{\delta ^{2}}{16}\text{, }h_{1}=-\frac{3}{8}\delta ^{2},\text{ }%
h(y_{0})=-\frac{\delta ^{2}}{2}\text{.}  \label{hh}
\end{equation}%
(For $\beta >0$, one can obtain the equation $6y-h_{1}=-3\delta \sqrt{y}$
but in combination with (\ref{h1}) it would give $y<0$ that is unacceptable
since outside the horizon we should have $y>0$. Thus this case should be
rejected.)

Then, using (\ref{xe}), (\ref{hl}), (\ref{om}) we find the results (\ref{be0}%
) - (\ref{nu}).


\begin{thebibliography}{99}
\bibitem{rufk} D. Pugliese, H. Quevedo, and R. Ruffini, Phys. Rev. D \textbf{%
84} 044030 (2011) (2013) [arXiv:1105.2959].

\bibitem{ruf3} D. Pugliese, H. Quevedo, and R. Ruffini, Phys. Rev. D \textbf{%
88} 024042 (2013) (2013) [arXiv:1303.6250].

\bibitem{ac1} N. I. Shakura and R. A. Sunyaev, Astron. Astrophys. \textbf{24}%
, 337 (1973).

\bibitem{ac2} Don N. Page and Kip S. Thorne, Astrophys. J. \textbf{191}, 499
(1974).

\bibitem{72} J. M. Bardeen, W. H. Press, and S. A. Teukolsky, Astrophys. J. 
\textbf{178}, 347 (1972).

\bibitem{ted11} T. Jacobson Class. Quantum Grav. \textbf{28} 187001 (2011)
[arXiv:1107.5081].

\bibitem{ind2} P. Pradhan and P. Majumdar, Eur. Phys. J. C \textbf{73}, 2470
(2013) [arXiv:1108.2333].

\bibitem{m} S. Ulbricht and R. Meinel, Classical Quantum Gravity 32, 147001
(2015) [arXiv:1503.01973].

\bibitem{extc} O. B. Zaslavskii, Phys. Rev. D 92, 044017 (2015)
[arXiv:1506.00148].

\bibitem{ban} M. Ba\~{n}ados, J. Silk and S.M. West, Phys. Rev. Lett. 
\textbf{103} (2009) 111102 [arXiv:0909.0169].

\bibitem{jl} O. B. Zaslavskii, Pis'ma ZhETF \textbf{92}, 635 (2010) (JETP
Letters \textbf{9}2, 571 (2010)), [arXiv:1007.4598].

\bibitem{fr} V. P. Frolov, Phys. Rev. D\textbf{\ 85}, 024020 (2012)
[arXiv:1110.6274].

\bibitem{weak} T. Igata, T. Harada, and M. Kimura, Phys. Rev. D \textbf{85},
104028 (2012) [arXiv:1202.4859].

\bibitem{circkerr} T. Harada and M. Kimura, Phys. Rev. D \textbf{83} 024002
(2011) [arXiv:1010.0962].

\bibitem{circ} O. B. Zaslavskii, Class. Quant. Grav. \textbf{29 }205004
(2012) [arXiv:1201.5351].

\bibitem{k} O. B. Zaslavskii, Phys. Rev. D \textbf{84}, 024007 (2011)
[arXiv:1104.4802].

\bibitem{vis} M. Visser, Phys.Rev.D 46 2445 (1992) [arXiv:hep-th/9203057].

\bibitem{string} A.A. Tursunov, \ M. Kolo\v{s}, A.A. Abdujabbarov, B.J.
Ahmedov, and Z. Stuchl\'{\i}k, Phys. Rev. D \textbf{88}, 124001 (2013)
[arXiv:1311.1751].

\bibitem{mpl} O. B. Zaslavskii, Mod. Phys. Lett. A \textbf{29}, 1450112
(2014) [arXiv:1403.6286].

\bibitem{magk} O. B. Zaslavskii, Mod. Phys. Lett. A \textbf{30}, 1550027
(2015) [arXiv:1407.5440].

\bibitem{tur2} A.A. Abdujabbarov, A.A. Tursunov, B.J. Ahmedov and A.
Kuvatov, Astrophys. Space Sci. \textbf{343}, 173 (2012).

\bibitem{wald} R. M. Wald, Phys. Rev. D \textbf{10}, 1680 (1974).

\bibitem{bron} K.A. Bronnikov and O.B. Zaslavskii, Phys.Rev. D \textbf{78}
021501, 2008 [arXiv:0801.0889]; K.A. Bronnikov and O.B. Zaslavskii, Class.
Quant.Grav.\textbf{26 }165004, 2009 [arXiv:0904.4904].

\bibitem{ag} A. N. Aliev and D.V. Gal'tsov, Sov. Phys. Usp. \textbf{32}, 75
(1989).

\bibitem{fs} V. P. Frolov and A. A. Shoom, Phys. Rev. D\textbf{\ 82}, 084034
(2010) [arXiv:1008.2985].

\bibitem{piat} A. A. Grib, Yu. V. Pavlov and O. F. Piattella, Int. J. Mod.
Phys. A \textbf{26} (2011) 3856 [arXiv:1105.1540].

\bibitem{ernst} F. J. Ernst, Journ. of Math. Phys., \textbf{17}, 54 (1976).

\bibitem{galpet} D.V. Gal'tsov and V. I. Petukhov, Sov. Phys. JETP \textbf{47%
}, 419 (1978).

\bibitem{ek} F. J. Ernst and W.\ J. Wild, Journ. of Math. Phys., \textbf{17}%
, 182 (1976).
\end{thebibliography}
\end{document}